\documentstyle[preprint,aps,pra]{revtex}

\def\bra#1{\langle #1 |}
\def\ket#1{| #1\rangle}

\begin{document}
\title{Against Many-Worlds Interpretations}
\author{Adrian Kent\thanks{Present address: DAMTP, University of
    Cambridge, Silver Street, Cambridge CB3 9EW, U.K.\hfill\break
Email: apak@damtp.cam.ac.uk}}
\address{The Institute for Advanced Study, School of Natural Sciences,\\
Princeton, New Jersey 08540, USA}
\maketitle
\begin{abstract}
\noindent
This is a critical review of the literature on many-worlds interpretations 
(MWI), with arguments drawn partly from earlier critiques by Bell and Stein.
The essential postulates involved in various MWI
are extracted, and their consistency with the evident physical world is
examined. 
Arguments are presented against MWI proposed by Everett, Graham and DeWitt.
The relevance of frequency operators to MWI is examined; it is argued
that frequency operator theorems of Hartle and
Farhi-Goldstone-Gutmann do not in themselves provide a probability 
interpretation for quantum mechanics, and thus neither support existing MWI 
nor would be useful in constructing new MWI. 
Comments are made on papers by Geroch and Deutsch that advocate MWI.
It is concluded that no plausible set of axioms exists for an MWI that
describes known physics.
\vfill
Int. J. Mod. Phys. A {\bf 5} 1745-1762 (1990)
\end{abstract} 
\eject
\centerline{\bf Foreword to archive version}

Though this paper was written in 1989, I thought it worth 
putting it on the archive now, since the debate to which it contributes is 
still very much alive --- and since some of the misconceptions 
which it addresses are still, I fear, a source of confusion.   

There is clearly enormous appeal in the idea that, somehow, there 
simply {\it cannot} be any real problem in extending 
quantum theory from the Copenhagen interpretation to a 
fully satisfactory interpretation of quantum cosmology. 
It seems to me that faith in that idea now generally survives 
despite, rather than because of, the many different attempts by 
distinguished physicists to explain precisely how the quantum
theory of the universe should be interpreted.  

This is not to dismiss the work of those pioneers. 
Even the harshest critic would accept that the many-worlds 
literature, and the work of the various post-Everettian schools,
includes a series of well-motivated and ingenious attempts 
to solve a problem which lies at the heart of modern physics.  
Whether or not those attempts have succeeded, they have certainly 
taught us more about the nature of the problem.  

And anyone who discusses these problems with colleagues quickly 
realises that there are thoughtful people who have considered
the counter-arguments and still believe 
that the problem was essentially sorted out, once and for all, 
by Everett [F1].  There are, similarly, thoughtful people who believe 
it has been resolved by DeWitt [F2], by Deutsch [F3],
by Coleman's exposition [F4] of Farhi--Goldstone--Gutmann [F5], by
Gell-Mann and Hartle's consistent histories approach to quantum
cosmology [F6], and so on.  

My impression, though, is that each of these camps represents a 
relatively small minority.  When any given attempt at something
like a many-worlds interpretation is spelled out and analysed, its 
problems turn out to be sufficiently serious to cause most 
physicists to look elsewhere.  Those who remain adherents 
tend to have non-standard views on the nature of scientific 
theories.    There are, for example, hard-line Everettians who
take Bell's intentionally pathological Bohm-Everett hybrid [F7] as
the inspiration for a serious proposal --- and happily accept 
that, when quantum cosmology is properly understood, 
it becomes clear that past events generally have no correlation 
with our present observations. 
Similarly, there are consistent historians who accept that no 
useful scientific predictions can be derived from their formalism
in the absence of a set selection hypothesis, and yet argue that
there is no need for such a hypothesis. 

Most people attracted by many-worlds ideas, however,
tend to be working in other areas of physics.
My impression is that most of these enthusiasts tend not to
follow very closely the debates as to what, precisely, any
given interpretation means and what its scientific implications
are.  And understandably so.  The basic idea of a many-worlds
interpretation seems superficially attractive: it is always 
unsettling to think that there may be an underlying 
problem which, at some stage, will have to be confronted, and 
it is reassuring to be told that there is an apparently
unthreatening solution.  Advocates of many-worlds interpretations 
tend, naturally, to stress the interest in their ideas, rather
than the problems --- which, of course, may not {\it be} such
serious problems from their perspective.  And it is hard to 
pick one's way through the literature on interpretations of 
quantum theory, and easy to come away with the (unfortunately 
sometimes correct) impression that nothing very constructive 
comes of these debates. 

It is really that audience --- those attracted to the idea of 
a many-worlds interpretation, but not professionally committed
to some particular line of thought --- that the paper was, and still
is, mainly meant to address.  

I have left the published version unaltered, except that  
references to preprints have been replaced by references
to the published papers.  The most obvious omission, in
hindsight, was any discussion of many-worlds-type interpretations 
based on the Schmidt decomposition, and more generally any 
discussion of that school of thought which suggests that 
the interpretational problems of quantum theory can be 
completely resolved by a proper understanding of the physics of 
decoherence.  A recent attempt to develop these ideas,
and to explain the obstacles that apparently prevent them
from succeeding, can be found in Ref. [F8]

A similar paper written today would also cover the 
consistent histories approach to quantum mechanics 
developed in recent years by Griffiths and Omn\`es and,
in the context of quantum cosmology, by Gell-Mann and Hartle.
As noted above, that approach too has run into serious problems, 
discussed in some detail in Refs. [F9-F13]. 

\centerline{\bf Foreword references}

\begin{description}
\item{[F1]} Refs. [1] and [19] below. 
\item{[F2]} Ref. [3] below 
\item{[F3]} D. Deutsch, Int. J. Theor. Phys. {\bf  24} 1 (1985); for a
critique see S. Foster and H. Brown, Int. J. Theor. Phys. {\bf 27}
1507 (1988).  See also Ref. [7] below. 
\item{[F4]} S. Coleman, {\it Quantum Mechanics In Your Face}, unpublished
lecture.
\item{[F5]} Ref. [5] below 
\item{[F6]} See e.g M. Gell-Mann and J.~B. Hartle,  in {\em Complexity, 
  Entropy and the Physics of   Information}, Vol.~III of {\em 
  SFI Studies in the Science of Complexity},   edited by 
  W.~H. Zurek (Addison Wesley, Reading, 1990).
\item{[F7]} Ref. [8] below
\item{[F8]} A. Kent and J. McElwaine, Phys. Rev. A {\bf 55} 1703 (1997). 
\item{[F9]} J.~P. Paz and W.~H. Zurek, Phys. Rev. D {\bf 48},  2728  (1993).
\item{[F10]} F. Dowker and A. Kent, Phys. Rev. Lett. {\bf 75} 3038 (1995). 
\item{[F11]} F. Dowker and A. Kent, J. Stat. Phys. {\bf 82} 1575 (1996).
\item{[F12]} A. Kent, Phys. Rev. A {\bf 54} 4670 (1996). 
\item{[F13]} A. Kent, gr-qc/9604012, to appear in Phys. Rev. Lett. 
\end{description}

\eject
\section{Introduction}

There is increasing interest in the foundational problems of quantum theory and
possible physical theories which might provide solutions.  
The subject is fascinating in itself, and is further motivated because
theorists continue to find that the barriers to understanding 
quantum gravity and cosmology are as much conceptual as calculational.  
Many-worlds interpretations (MWI) are often invoked.  Much theoretical work now
ultimately depends on the assumption that MWI are consistent with the
evident physical world.  Clearly, this consistency needs to be
examined.

The useful criticisms of MWI do not stem from an inability to accept
the picture of multiple branching universes, nor do serious critics
merely say that they do not see how to reduce macroscopic physics to
an MWI.  They assert either that particular physical facts are
demonstrably not logically deducible from the axioms of MWI, or else
they criticize the axioms proposed (usually on grounds of complexity
or arbitrariness).  Though it does not seem to be widely recognized,
these criticisms have had a substantial effect: modern opponents of
state vector reduction, recognizing that Everett's original argument
is erroneous, have adopted increasingly sophisticated formulations.

A popular misconception should be removed.  Many physicists have the
impression that the content of an MWI is {\em simply} the statement
that physical correlations can be calculated without state vector
reduction.  This statement is certainly consistent with known
experiments.  It may also be useful in simplifying the axioms of an
ensemble formulation of quantum mechanics.  
However, in this setting the statement does not
make a direct connection between the mathematical formalism and
physical reality (in particular, it actually says nothing about the
existence of many worlds).  The proponents of MWI were
motivated by greater ambitions; they aimed to produce a simple
mathematical theory whose formalism had a well-defined correspondence
with physical reality.  It will be argued that they did not succeed.

Even in modern formulations MWI still seem
seriously inadequate.
Thus the aim of this paper is to argue that the literature neither contains 
nor suggests a plausible set of axioms for an MWI that describes
known physics. 
Specifically, arguments are presented against
MWI proposed by Everett,\cite{ev} Graham\cite{gr} and
DeWitt;\cite{dew} and against the possibility of basing MWI on the
work of Hartle\cite{har} and Farhi-Goldstone-Gutmann.\cite{fgg} Some
brief comments, in reply to arguments put forward by Geroch\cite{ger}
and Deutsch\cite{deutsch} in favour of MWI, are also included.

The classic criticisms of MWI are Bell's papers.\cite{bell1,bell2}
Other published arguments against MWI include papers by
Stein\cite{stein} and Healey,\cite{healey} 
and comments by Shimony\cite{shim} and Wheeler.\cite{wheel}
The arguments below against Everett's and DeWitt's MWI 
draw largely from Bell's discussions, and also make use of 
Graham's criticism\cite{gr} of Everett's original argument;
some of Stein's arguments have also been adopted. 
Still, we hope that this article may be useful for two reasons.  
Firstly, we have tried to give a complete and self-contained criticism of 
all the ideas currently invoked in support of MWI. 
In particular, we have tried to refute claims that a probability 
interpretation for MWI can be defined by Graham's arguments about 
world-counting, or by the work of Hartle and Farhi-Goldstone-Gutmann on
the properties of frequency operators.  
Secondly, we have tried to clarify the logical structure of the MWI,
and so to pinpoint their essential problems. 
This involves the extraction and formalization of what seem to us the
fundamental assumptions in the original papers.  
The attempt may not have entirely succeeded.
But we are convinced that the {\em procedure} is justified, 
and in fact that axiomatization should have been insisted upon 
from the beginning. 
For any MWI worth the attention of physicists must surely be a
physical theory reducible to a few definite laws, not a
philosophical position irreducibly described by several pages of
prose. Moreover, once an axiomatization is achieved,
the irrelevance of much of the rhetorical argument surrounding MWI is exposed, 
and any given MWI's (de)merits become much clearer. 

In seeking this sort of clarification, we are not asking for anything 
extraordinary.  Simply, MWI proponents need to specify precisely the
mathematics involved in setting up their theory, to state 
explicitly which of the quantities they define are supposed to be 
physical, and to explain (in principle, not necessarily in detail) how our 
observations are
to be described in terms of these physical quantities.  
Examples of theories which satisfy these criteria are the standard 
presentations of Maxwell's electrodynamics and general relativity.
Here the electric and magnetic fields, matter and charge
densities, space-time manifold and metric are taken to be physical quantities,
while the gauge potentials and the various coordinate systems on the manifold
are only part of the mathematical formalism. 
(A more profound discussion can be found in \cite{bell4}.) 
Given standard (e.g. SI) definitions of mass, length, time, charge, 
current, and so forth, we understand how to identify the physical quantities
with the actual objects in our experience --- for example, we know how
to make a good approximation to a spherical test body of a given mass, and to
check that when put into orbit it does very closely follow what we calculate
to be a geodesic. 
(Of course, we really know none of these things with absolute rigour --- we are
dealing with physics, not pure mathematics, and we tacitly invoke many 
assumptions whenever we interpret an experiment (that scientific induction and
statistical inferences are valid, that no untoward background signal affects 
the data, that our carefully machined test mass remains internally stable and 
homogeneous while being transported to the laboratory --- we certainly cannot 
justify this from Maxwell's and Einstein's theories, \ldots). 
But, since it is claimed that in a many-worlds interpretation 
``the symbols
of quantum mechanics represent reality just as much as do those of 
classical mechanics,''\cite{dew} we should at least demand that an MWI 
be defined to the same level of rigour that classical theories are.) 

The dominant mode of research in current physics emphasizes mathematical
structure over problems of interpretation. 
So perhaps it needs to be said that we see this paper (like those it 
criticizes) as a discussion of physics, not philosophy. 
Of course it has often been a useful tactic to pursue calculations without
worrying much about their exact interpretation: the development of quantum
field theory is an outstanding example.
But, ultimately, we need not just to calculate a theory's predictions for 
familiar types of experiment, but to understand the theory's properties. 
The reason is that this is crucial in deciding the 
extent to which a theory has 
been tested, and hence in designing new tests.  
For example, once it was shown that standard quantum mechanics predicts 
non-local correlations,\cite{bell0} it became clear that new experiments
were needed to distinguish between the standard theory and local 
hidden-variable theories. 
Similarly, the physical community ought to be able to, and ought to, 
decide whether any many-worlds interpretation of quantum mechanics gives a 
precisely formulated mathematical model in which particular specified 
quantities have a well-defined correspondence to physics. 
For if not (and if the same conclusion is reached about other standard
interpretations), then it will presumably be generally agreed that 
precise models ought to be built, and tests designed
to distinguish these models from the standard theory. 
(A prototype model has already been 
suggested by Ghirardi-Rimini-Weber;\cite{grw} it is 
briefly discussed at the end of this paper.) 

After this introduction, the argument proper begins with a brief
explanation of what is meant by an MWI.  Here there is apparently no
controversy in the literature; very similar explanations have been
given by proponents of MWI, for example in the introduction to
reference \cite{dew}.  Next the papers by Everett, DeWitt and Graham
are considered in turn.  This is followed by a discussion of the
physical relevance of frequency operator theorems due to Hartle and
Farhi-Goldstone-Gutmann; the main claim here is that such theorems
neither support, nor aid in the construction of, MWI.  Some comments
on arguments by Geroch and Deutsch are then offered.  A concluding
discussion summarizes the problems with MWI and briefly considers
alternatives.

\section{The Case Against}

Many-worlds interpretations share two essential characteristics.
First, they suppose that there exists a definite physical reality,
which can be put into correspondence with parts of a mathematical
formalism.  This assumption is necessary if an MWI is to have any
useful content. It is an assumption made by Everett:\cite{ev} ``This paper
\ldots postulates that a wave function \ldots supplies a complete
matehmatical model for every \ldots system without exception.'' and
still more clearly by DeWitt:\cite{dew} ``The real world \ldots
is faithfully represented solely by the following collection of 
mathematical objects. \ldots The use of this word [faithfully] implies
a return to naive realism and the old-fashioned idea that there can be a 
direct correspondence between formalism and reality. \ldots The symbols
of quantum mechanics represent reality just as much as do those of 
classical mechanics.''\footnote{Actually, DeWitt here seems 
unnecessarily defensive; mathematical realism need not be naive, and has more 
or less continuously been advocated by distinguished physicists 
throughout this century.}
There seems to be no dispute in the literature on this point: if a theory
is not mathematically realist then it is not an MWI.

Second, they base the mathematical formalism on a state-vector
$\ket{\psi}$ which belongs to a Hilbert space $V$ and has a purely
hamiltonian evolution.  Throughout we shall use the Schr\"{o}dinger
picture and consider non-relativistic quantum mechanics.  Thus we can
formalize the fundamental mathematical assumptions as:
\begin{description}
\item[Axiom $0$]  There exists a Hilbert space $V$, a hermitian operator $H$ on
$V$, and a continuum of states $\ket{ \psi (t) } \in V$ for $-\infty <
t < \infty$ such that
\begin{equation}\label{schrod}
H \ket{ \psi (t) } = i \hbar \frac{ \partial }{\partial t} \ket{ \psi
(t) }~~
\end{equation}
holds for all $t$.
\end{description}

Different MWI vary in the extra formalism they introduce and in the
mathematical quantities which they identify as corresponding to
reality.  However, any meaningful MWI must include mathematical axioms
defining the formalism and physical axioms explaining what elements of
the formalism correspond to aspects of reality.  Now let us consider
specific cases.

\subsection{Everett}

Everett's famous paper\cite{ev} is admirably clear in its claims. 
For Everett, the mathematical formalism is fully described by Axiom $0$. 
Physical reality is to be entirely encapsulated in the wave function 
evolution. 
Everett is quite explicit in rejecting any further interpretational
axioms: ``The wave function is taken as the basic physical entity 
with {\em no} a priori interpretation.''
Thus we have only Axiom $0$ and:
\begin{description}
\item[Axiom 1E] The graph of the state vector's evolution 
(that is, the set of coordinates
$(\ket {\psi (t)},t) \in V \otimes (-\infty , \infty)$)
is a physical quantity. 
\end{description}

Now Everett points out that 
the Hilbert space inner product defines a measure $\mu$, by setting the measure
of a state \mbox{$\phi = \sum_{i} a_i \phi_i$} to be \mbox{$\mu( \phi ) = 
(\sum_i | a_i |^2 )^{\frac{1}{2}}$}, where the $\phi_i $ form an orthonormal 
basis of $H$. The measure has the property that if $\psi = \sum_j \psi_j$ and
the $\psi_j$ are orthogonal, then $\mu (\psi ) = \sum_j \mu (\psi_j )$. 
This measure $\mu$ is defined by the Hilbert space inner product, and so
by Axiom $0$ is part of Everett's mathematical formalism.

However, $\mu$ does not appear in axiom $1$E, and 
so is not a {\em fundamental} 
quantity in Everett's physical interpretation.
Nor does Everett make any attempt to show that $\mu$ can be understood as
a {\em derived} quantity in the physical interpretation. 
This leaves no way to deduce any statement connecting $\mu$ with real physics.
Since $\mu$ is of course precisely what we need to describe the measurement
correlations predicted by quantum mechanics, Everett's MWI is 
inadequate. 

The problem is illustrated by a simple 
example.\footnote{This example was apparently
first considered by Graham;\cite{gr} the 
following analysis combines his criticisms with those of Bell.\cite{bell2}
Anandan\cite{anandan} has also briefly mentioned the example, making the 
parenthetical point of the third criticism below, which he attributes to 
Aharonov.} 
Suppose that a previously polarized 
spin-$\frac{1}{2}$ particle has just had its 
spin measured by
a macroscopic Stern-Gerlach device, on an axis chosen so that 
the probability of measuring spin $+\frac{1}{2}$ is $\frac{2}{3}$. 
The result can be idealized by the wave function
\begin{equation}
\phi = a \, \phi_0 \otimes \Phi_0 + b \, \phi_1 \otimes \Phi_1
\end{equation}
where $\phi_1$ is the spin $+\frac{1}{2}$ state of 
the particle, $\Phi_1$ the state of the device having measured 
spin $+\frac{1}{2}$; $\phi_0$, $\Phi_0$ likewise correspond to 
spin $-\frac{1}{2}$; $|a|^2 = \frac{2}{3}$, $|b|^2 = \frac{1}{3}$.
Now in trying to interpret this result we encounter the following problems:

Firstly, no choice of basis has been specified; we could expand $\phi$ in 
the 1-dimensional basis $\{ \phi \}$ or any of the orthogonal
2-dimensional bases 
\begin{equation}
\{ \cos \theta \, \phi_0 \otimes \Phi_0  + 
\sin \theta \, \phi_1 \otimes \Phi_1 ,
 \sin \theta \,  \phi_0 \otimes \Phi_0   - 
\cos \theta \, \phi_1 \otimes \Phi_1 \}
\end{equation}
or indeed in multi-dimensional or unorthogonal bases. 
Of course, the information is in 
the wave function is basis-independent, and one is free to choose any
particular basis to work with.  But if one intends to make a physical
interpretation only in one particular basis, using quantities (such
as $|a|^2$ and $|b|^2$) which are defined by that basis, one needs to
define this process (and, in particular, the preferred basis) by an
axiom. This Everett fails to do.  

(A referee objects: ``As far as I understand it, the world is ultimately
described by quantum mechanics yet it appears classical to most macroscopic
observers. Phase information is effectively lost when measurements take
place. Exactly how this occurs clearly depends on dynamics. It is not going
to be part of a proper axiomatic formulation. \ldots Criticism of [an MWI]
because it does not, through its axioms, tell you how the branching takes
place is an example of wanting more from the axioms than they could ever 
reasonably provide.'' 

It's certainly true that phase information 
loss is a dynamical process which needs no
axiomatic formulation. However, this is irrelevant to our very simple point: 
no preferred basis can arise, from the
dynamics or from anything else, unless some basis selection rule is given. 
Of course, MWI proponents can try to frame such a rule in terms of 
a dynamical quantity --- for example, some measure of phase information loss.
But an explicit, precise rule is needed.)

Secondly, suppose that the basis 
$(\phi_0 \otimes \Phi_0 , \phi_1 \otimes \Phi_1 )$ is somehow selected. 
Then one can perhaps {\em intuitively} 
view the corresponding components of $\phi$ 
as describing a pair of independent worlds. 
But this intuitive interpretation goes beyond what the axioms
justify; the axioms say nothing about the existence of multiple physical
worlds corresponding to wave function components. 

Thirdly, in any case, no physical meaning has been 
attached to the constants $|a|^2$ and $|b|^2$.  
They are not to be interpreted as the probabilities that their respective
branches are realized; this is the whole point of Everett's proposal. 
It can not be said that a proportion $|a|^2$ of the total number of worlds
is in state $\phi_0 \otimes \Phi_0$; there is nothing in the axioms to 
justify this claim. 
(Note that if the two worlds picture {\em were} justified, then each state 
would correspond to one world, and it must be explained why each measurement 
does not have probability $\frac{1}{2}$.)
Nor can one argue that the probability of a particular observer finding
herself in the world with state $\phi_0 \otimes \Phi_0$ is $|a|^2$; 
this conclusion again is unsupported by the axioms. 

It is true that Everett attempts to relate the measure $\mu$ to physics by a
discussion of the memory of observers. 
But, given Everett's assumptions, the discussion
is actually free of physical content. 
Thus let $\Phi$ be the state vector of an idealized observer who has witnessed
$N$ measurements of systems identical to the one above; 
let $\Phi (i_1 , \ldots , i_N )$ describe the state of having witnessed
results $(i_1 , \ldots , i_N )$, where each $i_j$ is $1$ (or $0$) if the 
$j$-th observation was spin $+\frac{1}{2}$ (respectively 
$-\frac{1}{2}$); we shall 
suppress the correlated state vectors of the measured particles. 
We have the expansion 
$\Phi = \sum  a_{i_1 , \ldots , i_N}\Phi (i_1 , \ldots , i_N )$.
Everett considers the vector
\begin{equation}
\Phi^{\epsilon} = 
\sum_{| \frac{i_1 + \ldots i_N}{N} - \frac{2}{3} | > \epsilon}
 a_{i_1 , \ldots , i_N}\Phi (i_1 , \ldots , i_N )
\end{equation}
and shows that $\mu(\Phi^{\epsilon}) \rightarrow 0$ as $N \rightarrow \infty$
for any $\epsilon > 0$. 
But no new statement about physics can arise from this purely mathematical
derivation.  
Even supposing the infinite limit were obtained, which in reality is not
the case, the fact that
$\mu (\Phi^{\epsilon})=0 $ cannot imply that $\Phi^{\epsilon}$ is
physically irrelevant, when no hypothesis has been made about the
physical meaning of $\mu$.

(Reference \cite{mwbook} also contains a
longer exposition\cite{ev2} of Everett's ideas; this doesn't seem to depart 
from Everett's original paper on any point of principle.) 

\subsection{Graham}

The physical example discussed above is also helpful in understanding 
Graham's position.  
Graham criticizes Everett's argument, essentially for the reasons given in
the last section's penultimate paragraph. 
In fact, Graham interprets the wave function 
$\phi = a \, \phi_0 \otimes \Phi_0 + b \, \phi_1 \otimes \Phi_1$
as describing precisely two worlds, which (so long as $a$ and $b$ are non-zero)
are equiprobable. 
He claims that, if this were indeed the wave function after measurement,
then the probability of measuring 
spin $\pm \frac{1}{2}$ would be the naive ``counting probability'' of 
$\frac{1}{2}$.
However, he argues that in a more realistic analysis (in which the measuring 
device has many degrees of freedom and must be described by a statistical
mixture of quantum states) the counting probability of a particular
measurement can be shown to {\em equal} to the standard quantum mechanical
probability (defined by $\mu$). 
This is a remarkable claim, which has attracted strangely little scrutiny. 

Let us try to axiomatize Graham's position.  In addition to Axiom 0, we 
suggest:
\begin{description}
\item[Axiom 1G] A preferred orthogonal 
basis is defined for the universal wave function,
in which each basis element defines a state 
where all macroscopic measuring devices register definite results.  

\item[Axiom 2G]
Suppose that at time $t$ the wave function $\psi(t)$ has the expansion 
\begin{equation}
\psi(t) = \sum_{1 \leq i_1 \leq j_1 , \ldots , 1 \leq i_N \leq j_N} 
a_{i_1 , \ldots , i_N} \psi (i_1 , \ldots , i_N )
\end{equation}
where $\psi (i_1 , \ldots , i_N )$ is a state in which $N$ discrete
acts of measurement, whose possible ranges were $(1,2, \ldots j_r)$, 
obtained the results $i_r$ (where $r$ runs from $1$ to $N$). 
Then the physics corresponding to $\psi$ is that at time $t$ there 
exist independent classical worlds corresponding to each 
$\psi (i_1 , \ldots , i_N )$ for which $a_{i_1 , \ldots , i_N} \neq 0$
(i.e. a maximum of $\prod_{r=1}^N  j_r$ such worlds), in which the 
measuring devices have registered the results described 
by $\psi (i_1 , \ldots , i_N )$. 
\end{description}

Now these axioms are open to criticism on many grounds --- most obviously
those of vagueness and complexity. (Note that axiom 1G is an axiom defining
further mathematical formalism, yet involves a measuring device --- at best
a very unwieldy mathematical object.)
Ignoring these problems for the moment,
let us accept that, for any macroscopic observer, axiom 2G implies
probabilistic statements about the state of the observed world at time $t$:
she finds herself equiprobably in one the $M \leq \prod_{r=1}^N  j_r$ worlds,
and this defines the ``counting probability'' of any statement about 
the results of measurements in her world, at that particular instant. 
What we want to point out is that Graham's claim that the counting
probability equals the probability defined by the standard measure $\mu$ 
does {\em not} follow from his assumptions.

Graham's strategy is the following.  
He considers an observable $X$ with eigenvalues $X_i$ ($i$ from $1$ to $n$),
a large number $N >> n$ of identical systems on which $X$ is defined,
and a macroscopic measuring device that measures the 
{\em relative frequencies} of the various eigenvalues in the identical
systems. 
Finally he considers an observer examining the output of the measuring device.
Then he argues that, from statistical assumptions about the initial state
of the measuring device, it can be shown that after the second observation,
in nearly all worlds 
the relative frequencies observed will be very close to those predicted by
standard quantum mechanics. 

Now we do not think Graham satisfactorily establishes even this 
last, restricted claim. His argument relies on a quantum statistical 
mechanical assumption about the probability distribution for the initial
(normalized) state $\psi$ of the macroscopic measuring device.
What he assumes is that $\psi$ is constrained to lie in some finite
($n$-)dimensional space ${\cal H}$, and that the expansion
\begin{equation}
\psi = \sum_{j=1}^{n} <~\psi_j~,~\psi~> \psi_j
\end{equation}
of $\psi$ in an orthonormal basis $( \psi_j )$ for ${\cal H}$ defines 
a statistical measure for $\psi$ as follows. 
Graham sets 
$<~\psi_j~,~\psi~> = x_j + i y_j$, and takes $(x_1 , \ldots , x_n , y_1 ,
\ldots , y_n )$ as coordinates on the $(2n - 1)$-sphere $S^{2n-1}$. 
Now he defines the statistical measure $\lambda$ 
for any set I of states as the ratio
\begin{equation}\label{measure}
\lambda (I) = 
\frac{{\rm surface~ area~ of~ I~ on} ~S^{2n-1}}{{\rm surface~ area~ of}~ S^{2n-1}}
\end{equation} 
Now this might be a natural statistical measure if the probabilities of
events are ultimately based on the standard 
quantum measure $\mu$, but there seems no reason to believe that it 
would be the right statistical measure if the counting 
probabilities were fundamental in nature. 
Since the whole point is to derive $\mu$ from the counting probabilities,
the argument seems circular. 
(It is of course useless to try and justify equation (\ref{measure}) by
empirical data, since these data are supposed to follow from the 
fundamental assumptions.)

In any case, the proposition that Graham needs to establish is much
more general than the one he actually considers.  
Measuring devices are not typically coupled to the
relative frequency operator; it is more normal to record the results of
individual observations in sequence.\footnote{This point, in a related context,
is made by Stein\cite{stein} (pp. 641-2).}
Whether one wants to describe this as
a complex observation by a single measuring device or 
a sequence of observations of identical systems by different measuring devices
does not matter. 
The point is that 
Graham's
argument simply does not cover this common (in fact generic) physical 
situation.
In fact, it is trivial to see that the counting probability of a {\em general}
sequence of observations of independent observables by independent 
measuring devices could only equal the standard quantum probability
if the equality were to hold for a {\em single} measurement of a general 
observable.  
Now the equality does not hold for a single spin measurement, in the
simple model discussed at the end of the last section.
So Graham would have to argue that the quantum statistical mechanics of a true
Stern-Gerlach measuring apparatus produces the counting 
probabilities $\frac{1}{3}$ and $\frac{2}{3}$ for the measurement discussed,
and different counting probabilities in experiments where the electron had
a different spin polarization.
This is patently absurd, and certainly does not follow 
from Graham's assumptions about the statistical 
distribution of the measuring device's initial states. 
We conclude that counting probabilities are not relevant to physics. 

\subsection{DeWitt}

DeWitt begins by explicitly presenting a ``postulate of mathematical
content'' and a ``postulate of complexity''.
The latter states 
that ``The world is decomposable into systems and apparata.'', 
and we take this as meaning that DeWitt would accept 
Axiom 1G above.\footnote{Although DeWitt suggests his postulates are already 
implicit in Everett's paper,\cite{ev} 
we agree with Bell\cite{bell2} that
the postulate of complexity is against the spirit of Everett's 
prose.}
However, DeWitt's position on the physical interpretation of the wave
function is not clear to us. 
DeWitt refers to Graham's argument as an improvement on Everett's (see the
footnote on page 185 of reference \cite{mwbook}), but does not actually 
invoke the counting probability at all.
He makes no explicit postulate about the correspondence between the state
vector and multiple worlds or their content.
His discussion (pages 183-6 of reference \cite{mwbook}) of the memory
content of automata is mathematically more sophisticated than that of
Everett, but unsatisfactory for the same reason --- the mathematical
statement that a particular component of the state vector has measure
zero is not physically interpretable when no physical properties of
the measure have been postulated.

(We here have to digress a little, since DeWitt himself appears to make
and then counter a point similar to ours: ``The alert student may now
object that the above argument contains an element of circularity. In
order to derive the {\em physical} probability interpretation
$\ldots$, we have introduced a {\em nonphysical} probability concept,
namely that of the measure of a subspace in Hilbert space.''  DeWitt
goes on to note that measures are also used in statistical mechanics.
Now we are not sure that the alert student is making quite the same
objection as ours: our objection is not that a measure is introduced
in DeWitt's formalism; it is that its r\^{o}le in the interpretation is
never explained.  DeWitt may thus not be addressing our objection at
all.  But it {\em is} sometimes claimed in defence of MWI that there are 
analogous problems in some branches of statistical mechanics.
It does not seem clear that the problems really are analogous,
but if they were, this of course would be an argument 
that statistical physics is not well-founded, not an argument that  
MWI are valid.\footnote{Our rejection of this analogy is essentially 
isomorphic to Bell's discussion
of classical and quantum mechanics on page 125 of reference \cite{bell2}.})

Although DeWitt is unexplicit on the r\^{o}le of the measure, there is
a formalization which does not seem in conflict with his writings. 
This formalization is a version of one due to Bell.\cite{bell1,bell2}
Classical variables have replaced the de Broglie-Bohm variables
in Bell's formalization; we regard this as uglier, but it respects the
spirit of DeWitt's axiom of complexity. 

Suppose we adopt Axiom 0 and Axiom 1G above, together with:
\begin{description}
\item[Axiom 2D] At any time $t$ there 
is a continuum of independent classical worlds 
$W_s (t)$, labelled by a real parameter $0 \leq s \leq 1$.
Physics consists in a statement of the possible classical contents (recorded by
measuring devices) of the various worlds, and the (standard real number)
measure of the set of labels $s$ corresponding to worlds
in which each classical content arises. 

\item[Axiom 3D] Suppose that at time $t$ the 
wave function $\psi(t)$ has the expansion 
\begin{equation}
\psi(t) = \sum_{r_1 \ldots r_N  }  \psi (r_1 , \ldots , r_N )
\end{equation}
where $\psi (r_1 , \ldots , r_N )$ is a state in which $N$ acts of measurement
obtained the results $r_i$ (where $i$ runs from $1$ to $N$ and the variables
$r_i$ may be continuous or discrete). 
Then the physics corresponding to $\psi$ is that at time $t$, there exist 
classical worlds corresponding to each set of results $(r_1 , \ldots , r_N )$,
and the set of these worlds has measure $\mu (\psi (r_1 , \ldots , r_N ))$. 
\end{description}

Now it is possible that, with a sufficiently long and careful classification
of all known measuring devices, these axioms would become consistent
with known experimental results. 
(Zurek\cite{zurek} has given a careful discussion of how a preferred basis
might be defined by measuring devices.)
Our reasons for rejecting an MWI of this type --- which appear below for
completeness, but are identical to those
given by Bell\cite{bell1,bell2} --- are therefore not strictly
empirical.   
The first objection is that, 
precisely because this classification would be needed, the axioms 
should not be taken seriously as possible fundamental physical laws. 
The reductionist approach --- explaining physical phenomena in terms of 
simple, mathematically precise, quantities --- has been extraordinarily
successful in almost all areas of physics.  
It goes against everything we have learned about nature to propose a theory
in which complicated macroscopic objects, whose precise definition must 
ultimately be arbitrary, are fundamental quantities. 

A second objection is that this formulation
destroys our concept of history.  
An observer at time $t$ is in a world whose state (and thus, the observer's
memory) has a certain probability distribution; at time $t^{\prime}$ 
another world, part of a different distribution, might contain an observer with
qualitatively similar physical features.  
But there is no real connection between the observers, and their memory 
contents
need not overlap at all. 
The idea of a classical object pursuing a (more or less) definite trajectory
through space time has been lost, and we do not see that it can be recovered
without extensively amending the formalism. 
This is a severe loss; it makes the formalism much uglier; it implies
that human experience is a sequence of disconnected fictions.
But perhaps these are not devastating weaknesses for the non-relativistic 
theory, considered purely as physics. 
However, the formalism also seems irreconcilable with special relativity.
This is perhaps the strongest physical argument against having
distributions of worlds $W_s (t)$ that are independent for each $t$. 

As mentioned earlier, Bell's de Broglie-Bohm MWI avoids the objection
of complexity, by working with particle position variables instead
of classical variables. It may alternatively be possible to change the
MWI above so as to regain the notion of a classical trajectory.
The real problem for MWI advocates is to find a formalism which {\em both} 
avoids complex anti-reductionist axioms and allows classical trajectories
to be defined. 

This ends the first part of our discussion.  
Working from the papers advocating MWI, we have tried to extract axioms that 
are simultaneously well-defined, plausible as fundamental laws of nature,
and consistent with known physics.
We have failed; we do not think it is possible.

\subsection{Note on Cooper-Van Vechten}

As well as the papers already discussed, DeWitt and Graham's book 
``The Many-Worlds Interpretation of Quantum Mechanics''\cite{mwbook} 
contains a paper by Cooper and Van Vechten.\cite{cvv}  
In a footnote, Cooper and Van Vechten explain the difference between
their interpretation and Everett's.  
They don't advocate any sort of many-worlds picture, nor do they take  
the wave function to be a physical quantity; the physical quantities
in their interpretation are events occurring in a single world. 
An event is defined 
as ``the interaction of a system to be observed or prepared with another system
which can be put into a state that is irreversible for reasons of entropy'' ---
anything which is usually called a measurement is supposed to fall into this
category. The correlations between events are calculated using purely
hamiltonian evolution (no state vector reduction is postulated). 

Here we just note that this isn't a many-worlds interpretation; its 
problems are essentially those of some standard interpretations of quantum 
mechanics, and beyond our present scope. 

\subsection{Hartle and Farhi-Goldstone-Gutmann} 

In a famous paper, Hartle defined and analysed the properties of
the relative frequency operator which acts on the Hilbert space of an infinite
ensemble of identical systems, by treating the operator as a limit of 
frequency operators defined on finite ensemble Hilbert spaces.  
Farhi-Goldstone-Gutmann (FGG) 
have recently presented a new treatment of relative
frequency, in which the operator is defined directly (i.e. without a 
limiting process) on states representing the infinite ensemble.   
Formal theorems about the infinite ensemble relative frequency operator 
are clearly regarded by proponents of MWI
as proving something useful to their case (Hartle's work is almost always
cited by post-$1968$ proponents), but it is hard to extract any definite
claims about precisely what use these theorems are 
{\em in the context of MWI.} (Clearly the theorems {\em are} useful in 
reformulating, or examining the foundations of, standard quantum mechanics.)
So it is worth looking at the actual implications of relative frequency
theorems for MWI. 
We examine FGG's results; aficionados of Hartle's should find it
easy to translate.
It should be stressed that references \cite{har} and \cite{fgg} do not 
directly advocate MWI; we make no criticism of reference \cite{har};
a criticism of the physical interpretation which FGG ascribe to their
results appears below; however, our main point in this section is to
argue against the relevance of frequency operators to MWI, and in this
argument we criticize proposals that are not suggested in either
paper.

FGG examine the consequences of the following postulates:
\begin{description}
\item[Postulate I] The states of a quantum system, S, are described by vectors
$\ket{\psi}$ which are elements of a Hilbert space, V, that describes
S.
\item[Postulate II] The states evolve in time according to
\begin{equation}
H \ket{\psi} = i \hbar \frac{ \partial }{\partial t} \ket{\psi}
\end{equation}
where H is a hermitian operator which specifies the dynamics of the
system S.
\item[Postulate III]
Every observable, ${\cal O}$, is associated with a hermitian operator
$\theta$.  The only possible outcome of a measurement of ${\cal O}$ is
an eigenvalue $\theta_i$ of $\theta$.
\item[Postulate I${\bf V}^{\prime}$]
If a quantum system is described by the eigenstate $\ket{\theta_i}$ of
$\theta$ (where \mbox{$\theta \ket{\theta_i} = \theta_i \ket{\theta_i}$)}
then a measurement of ${\cal O}$ will yield the value $\theta_i$.
\end{description} 
These differ from the standard postulates for quantum mechanics in
that postulate I${\rm V}^{\prime}$ has replaced the standard
postulate:
\begin{description}
\item[Postulate IV] If the state of the system is 
described by the normalized vector $\ket{\psi}$, then a measurement of
${\cal O}$ will yield the value $\theta_i$ with probability $p_i = |
<~\theta_i ~,~ \psi ~> |^2$.
\end{description} 
and in omitting the projection (or state vector reduction) postulate:
\begin{description}
\item[Postulate V]
Immediately after a measurement which yields the value $\theta_i$ the
state of the system is described by $\ket{\theta_i}$.
\end{description} 

The omission of postulate V does not mean that FGG are directly
arguing for pure hamiltonian evolution.  Their primary concern is to
analyse the probabilistic interpretation of a single act of
measurement, and their discussion largely ignores the question of
whether this measurement reduces the state vector.  (In the final
paragraphs of reference \cite{fgg}, FGG consider whether postulate V
is in fact necessary.  They claim that it can be neglected if one
introduces a definition of measuring device into the postulates,
although they do not necessarily advocate this procedure.  This view
possibly hints at an MWI rather like the formalization presented in
the above discussion of DeWitt's paper.  However, FGG's discussion of
how postulate V might be omitted is not very explicit; it is clearly
not the main point of their paper, and we shall not try to engage with
them on this question.)

In order to justify replacing postulate IV by postulate I${\rm
V}^{\prime}$, FGG consider the implications of postulates I-III,
I${\rm V}^{\prime}$ for an infinite ensemble of identical systems.
Let $S$ be the single system.  For simplicity, the state space of $S$
is taken to be two-dimensional.  Let ${\cal O}$ be an observable defined
on $S$, $\theta$ be the corresponding hermitian operator with
eigenstates $\ket {\theta_i}$ and eigenvalues $\theta_i$, so that
$\theta \ket {\theta_i} = \theta_i \ket{\theta_i}$ (for $i = 1$ or
$2$).  Now let S be the system comprising a (countably) infinite
number of copies of $S$, and let $\ket{\psi}$ be some state in $S$.
FGG show that
\begin{equation}\label{eigen}
F( \theta_i ) \ket{\psi}^{\infty} = p_i \ket{\psi}^{\infty}
\end{equation}
where the vector $\ket{\psi}^{\infty}$ represents the infinite
ensemble of systems all in the state $\ket{\psi}$, the frequency
operator $F( \theta_i )$ is associated with the observable that
describes the proportion of subsystems in the eigenstate
$\ket{\theta_i}$, and $p_i = | <~\theta_i ~,~ \psi ~> |^2$.

In summary, one knows from empirical evidence that the probability of
obtaining the result $\theta_i$ from a measurement of ${\em O}$ on the
state $\ket{\psi}$ is $p_i$.  FGG have clearly shown that this number
can in principle be {\em calculated} from postulates I-III, I${\rm V}^{\prime}$
without using the Hilbert space norm.  This is a new and interesting
result in the foundations of quantum mechanics.  However, FGG seem to
make a stronger claim (in section X of reference \cite{fgg}, and also
implicitly in its title: ``How Probability Arises in Quantum
Mechanics'').  This claim is that their formalism implies the physical
interpretation of (not merely a way to calculate) the number $p_i$, as
the probability that a measurement on a single system will obtain the
result $\theta_i$.

The latter claim does not seem to be right.  It is clear that (unless
$\ket{\psi}$ is an eigenstate of the operator $\theta$) postulate
I${\rm V}^{\prime}$ says nothing about measurements of $\theta$ on a
single system, or of the associated frequency operators on any finite
ensemble of systems.  The postulate applies only if the single system
is part of an infinite ensemble, and then it makes a statement about
the quantum state of the entire ensemble.  This implies no
probabilistic statement about the states of any of the ensemble's
members.  For the state $\ket{\psi}^{\infty}$ does not ascribe a
definite eigenstate to any individual system in the ensemble.  (It can
loosely be thought of as a superposition of all states in which the
systems are in definite eigenstates, of which a proportion $p_i$ are
$\ket{\theta_i}$.)

A famously similar situation arises in a $2$-slit diffraction
experiment.  Suppose we repeat many times an experiment in which a
single electron is directed towards a scintillating screen in which
there are two slits, beyond which there is a second parallel
unblemished screen, and consider one such experiment in which at 
the time when the electron passes the
plane of the first screen no flash is produced.  Then if
$\ket{\psi_1}$ and $\ket{\psi_2}$ are the wave functions associated
with an electron going through slits $1$ and $2$ respectively, the
electron's state after passing the first screen is $\ket{\psi_1} +
\ket{\psi_2}$.  This is an eigenstate of the hermitian operator
$\ket{\psi_1} \bra{\psi_1} + \ket{\psi_2} \bra{\psi_2}$ with
eigenvalue $1$.  There is a sense in which the statement ``the
electron went through precisely one slit'' would thus follow from
FGG's postulates.  But one has to be careful in translating the
mathematics into colloquial language.  In the sense in which the
statement follows from the mathematics, it is not a statement about
classical events; it therefore does not imply the statement ``the
electron went through slit $1$ with probability $\frac{1}{2}$''.
(This is of course good, since the last statement is inconsistent with
the cumulative diffraction pattern observed at the second screen).

We are not criticizing the standard use of infinite
ensembles to define probabilities. If it could be shown, {\em in the ordinary 
meaning of the words}, that a proportion $p_i$ of the infinite 
ensemble will be in the state $\ket{\theta_i}$, we should agree that a 
probability interpretation is established. 
But the qualification is crucial. The standard 
calculus of probabilities applies to sets of 
definite, classical events. It doesn't apply, or at least not 
straightforwardly, to quantum states which describe superpositions of many
different sets of such events. 
The statement ``the state $\ket{\psi}^{\infty}$ is an eigenstate of the 
relative frequency operator with eigenvalue $p_i$''
is weaker than ``the systems in the ensemble are each
in definite eigenstates of $\theta$, with a proportion $p_i$ in the 
eigenstate $\ket{\theta_i}$''. 
FGG establish the first statement, and then deduce a probability interpretation
from the second. We see no way to bridge this logical gap in their argument. 

So we have concluded that FGG's relative frequency theorem 
does not {\em per se} imply a probability interpretation 
for any individual system.  How then could it be 
used to construct an MWI?  There seem to be 
two possible lines of argument.  The first is to 
suppose that in reality our universe {\em is} part 
of an infinite ensemble of identical universes, so that 
the FGG theorem would apply directly.  The second would 
not require an infinite ensemble.  Instead it would use 
Hartle's observation\cite{har} that frequency operator theorems apply not only 
to ensembles, but also when a single system is 
found over time to be in the same state 
infinitely often.  (Whether any given system will return to 
the same state infinitely often is a complicated question 
in nonrelativistic quantum theory.  To answer the question in 
reality would seem to require both a quantum theory 
of gravity and a description of the initial state 
of the universe.  Let us concede that it might 
be true.)  The problem with both lines of argument 
is that they would describe physics as seen by 
supernatural observers who directly apprehend (or perhaps explicitly measure,
by the sort of methods discussed in section X 
of reference \cite{fgg}) the eigenvalues of frequency operators.
Unless the postulates were amended so as to specify 
what is meant by a measurement and a measuring 
device, these observers would receive data corresponding to the 
frequency operator eigenvalues of every possible observable; the first 
type of observer would receive these data at every 
instant in time, the second presumably at $t= \infty$.  In 
any case, the supernatural observers would have an 
entirely deterministic impression of physics.  Suppose the universe's wave 
function describes (at least in some components) observers of 
the ordinary type (humans or automata).  Then the supernatural 
observers would receive a frequency operator analysis (summed over 
ensembles or over time, assuming in the latter case 
that the state vector of a given observer will 
faithfully be reproduced infinitely often 
between $t = \pm \infty$) of these 
ordinary observers' observations and memories.  But now 
we have again reached the problem of how to 
extract a probabilistic interpretation, albeit in a different guise.
The supernatural observer's interpretations make no statements about the 
memories or observations of individual ordinary observers.  In particular,
if we are assumed to be the ordinary observers,
they do not account for our perception that definite 
events occur according to probabilistic laws.  

We conclude that relative frequency theorems do not
seem useful in 
defining MWI, nor do they seem relevant in 
deciding whether a given MWI is valid. 

\subsection{Comments on Geroch and Deutsch}

Some interesting arguments have been put forward by Geroch,\cite{ger}
in an article intended ``to state the content of, and to some extent to
advocate'' Everett's interpretation.  
Deutsch\cite{deutsch} 
has presented thought experiments involving various types of ``quantum 
parallel processing'', which are intended to make state vector reduction 
implausible.  We comment briefly on some points raised in these papers.

For Geroch, the notion of ``preclusion'' is fundamental.  Roughly speaking,
a region of configuration space is said to be {\em precluded} if the wave
function $\psi$ becomes at some time ``small'' in the region. 
The notion is illustrated by physical examples. 
Geroch then claims that: ``Every physical prediction of quantum mechanics can
be reformulated as the statement that a certain region is precluded.'' and
that the essence (or ``weak form'') of the Everett interpretation lies in
adopting the language of preclusion for quantum mechanics.

Now there are two obvious objections to this program. 
Firstly, it begins from a false premise. 
There {\em are} physical predictions of 
standard quantum mechanics which cannot be reformulated in the language
of preclusion, for example ``If a particle's spin along the $x$ axis is
$\sigma_x = +\frac{1}{2}$ and a measurement is made of $\sigma_y$, the
definite results $\sigma_y = \pm \frac{1}{2}$ will be obtained with
probability $\frac{1}{2}$.'' 
(The postulates of standard quantum mechanics, as framed by
FGG, were given in the last section as PI-PV. The statement in question
follows directly from PIV.) 
Secondly, the notion of preclusion is actually a rather complicated and 
vaguely defined one. 
Stein has discussed this at some length in his reply\cite{stein} to Geroch; 
we briefly summarize (what we agree with Stein are) the problems.
Preclusion seems to involve
a preferred basis (defined by measuring devices, as 
discussed in section 2.3) in order to define precisely {\em which}
small components of the wave function are to be precluded. 
And then, beyond failing to explain how definite results are
perceived, preclusion does not explain how to
calculate the probabilities of any given result, or of finite sequences 
of measurements: in short, it evades the question of how a probabilistic
interpretation arises for the state vector at any given finite time. 
So it does not seem that preclusion helps in defining axioms for an MWI.
Certainly Geroch does not claim to formulate axioms
using preclusion, or even to give a precise definition of preclusion.  
The point is that such a program could not be carried through, or at least 
could not produce a good description of known physics.
But then surely preclusion cannot really be a fundamental notion? 
 
Both Geroch and Deutsch suggest that there is a useful analogy
between the interpretations of general relativity and MWI.
In general relativity, they point out, humans sweep out a world-tube.
They further note that 
humans do not simultaneously perceive events occurring at all points within
the tube, yet few physicists consider rejecting general relativity for
this reason. 
The intended implication seems to be 
that the interpretational problems of a given MWI are 
similar, and so should be regarded as similarly unthreatening to the MWI's 
status. 

However, the two theories are actually radically different.
We don't think there is any fundamental obstacle to qualitatively
modelling the {\em content} of successive states of 
human consciousness using matter 
distributions on a manifold. 
Although one does not know how to reconcile the actual structure of matter
with the concepts of general relativity, one could presumably
design automata with input devices by
using idealized compressible fluids. 
Whether this would produce a satisfactory model of human
consciousness or not, it surely wouldn't teach one 
how to obtain a probabilistic classical interpretation from 
a Hilbert space vector. 
Now the intended point of the analogy may be that general relativity has not
been shown to,
or perhaps can not, explain the {\em fact} of human consciousness.  
If so, the same response applies here as to earlier analogies: the problem
of defining or predicting self-awareness is a problem with general relativity;
it may be a problem with any mathematical theory of nature; making this
observation does not help to define an MWI. 
So, however intended, the analogy does not seem helpful.\footnote{The points 
made in this paragraph are greatly elaborated in 
pages 644-6 of reference \cite{stein}.}

Finally, we note that Deutsch's main discussion involves thought
experiments (Deutsch's experiments $2$ and $3$) whose outcome is quite
uncertain. Deutsch assumes that ``quantum parallel processing''
(which relies on pure hamiltonian evolution of the state vector) will occur
during the operation of various computing devices.  There is presently
no compelling reason for this assumption. 
(Nor would we necessarily interpret the results Deutsch predicts as compelling
evidence for MWI.)

\section{Discussion} 

Everett's original attempt at a many-worlds interpretation was an
interesting and well-motivated idea. 
Everett wanted simultaneously to simplify the axioms of quantum
mechanics and to produce a realist physical theory (one which 
gives an explicit mathematical description of an external reality). 
However, it is clear that he did not succeed. 
Successive advocates of many-worlds interpretations, while usually 
claiming merely to clarify Everett's position, have substantially abandoned 
his goals.  
No proponent of MWI (as far as we are aware)
has actually produced a complete set of axioms to
define their physical theory.  
It seems that any such axioms must involve the extremely complicated notion
of a measuring device, or that the physical reality they define will 
be one without any notion of timelike trajectory (and so, as 
Bell\cite{bell1,bell2} 
has pointed out, not obviously consistent with special relativity), or both. 

All of this suggests that Everett's original program has degenerated,  
apparently beyond repair.  
Although the logical possibility of an MWI consistent with known physics
cannot completely be excluded, it seems that the defining axioms
of such a theory would have to be extremely ugly and arbitrary.
How then can the continued attraction of MWI be accounted for?  
The question is really one for future historians of science,
but perhaps we can suggest two possible answers.

Firstly, the very failure of MWI proponents to axiomatize their
proposals seems to have left the actual complexity of realistic MWI widely
unappreciated.  
It may thus possibly be tempting for MWI advocates
to assume that there is no real problem;
that Everett's detractors
either have not understood the motivation for, or merely have rather
weak aesthetic objections to, his program.  (Hence perhaps the
otherwise inexplicable claim by one commentator that ``Avoiding this
[prediction of multiple co-existing consciousnesses for a single
observer] is their [Everett's opponents'] motivation for opposing
Everett in the first place.''\cite{deutsch})
Secondly, MWI seem to offer the attractive prospect of using quantum
theory to make cosmological predictions.  
The trouble here is that if an MWI is ultimately incoherent and ill-founded,
it is not clear why one should pay attention to
any quantum cosmological calculations based on it.
Perhaps the right attitude
should be to explore with interest, but to take any results that 
depend on MWI as no more than suggestions --- not to be taken seriously
without independent proof. 

Beyond the community of MWI theorists, Everett's legacy seems to be
the widespread belief that pure hamiltonian evolution is manifestly
the correct description of nature (and ``therefore'' --- although it
does not really follow --- that there is no problem with the foundations
of quantum mechanics and no need to consider alternative theories). 
Although we have not argued against the possibility of an acceptable physical
theory (other than an MWI)
involving pure hamiltonian evolution, many arguments against MWI still
apply. Most obviously, there is the problem of how a probabilistic 
interpretation with definite measurements arises from a deterministic
wave function evolution which superimposes the state vectors of all possible
measurements. 
So we do not see how to make such a theory precise without assumptions such as
those of de Broglie\cite{db} and Bohm,\cite{bohm} and are
not enthusiastic about the de Broglie-Bohm theory (for ultimately
aesthetic reasons --- the theory gives a definite 
physical interpretation for quantum mechanics,
but at the cost of introducing two levels of formalism related
in a mathematically uninteresting way; it suggests nature is
more malicious than subtle).  We suspect any better, realist theory
with purely hamiltonian evolution would have to augment or replace the
standard quantum formalism by radically new mathematical concepts.
In short, pure hamiltonian evolution seems a difficult hypothesis to
justify (unless perhaps one abandons realism --- and {\em that} seems
still less justifiable for a physicist), 
and certainty that it is correct seems 
scientifically incomprehensible.

Critics of MWI are often asked what alternative they propose.   
Of course the criticisms of MWI stand or fall
on their own merits, but let us briefly outline our best guess at the
actual status of quantum mechanics. 
Similar views have been expressed by Leggett\cite{leggett} and 
Penrose\cite{penrose} (among others) and in the references cited below. 
Our guess is that the postulates of standard quantum mechanics (I-V above)
are not an ultimate description of nature; the ultimate description is
given by postulates as yet unknown, which are probably framed in a
different mathematical language.
However, insofar as the language of state vectors in a Hilbert space applies
to physics, postulates I-V are a good approximation to the behaviour of 
microscopic systems. In particular, the state vector reduction postulate
is an approximate description of what happens to a microscopic
system when it interacts with a macroscopic one. 
For macroscopic systems, either the postulates do not
hold (the Schr\"{o}dinger equation ceases to be a good approximation and 
the postulate of state vector reduction ceases to be well-defined)
or the whole language of state vectors is inapplicable.  
The distinction between microscopic and macroscopic should be one of degree
not of kind, and should follow from the unknown fundamental postulates. 

Very interesting models of possible 
sub-quantum physics, apparently consistent with known experiment and
with all the properties required in
the last paragraph, now exist.  The key idea for these models was put
forward by Ghirardi-Rimini-Weber;\cite{grw} a very clear exposition and
new interpretation of the mathematics has been presented by Bell;\cite{bell3}
a possible treatment of indistinguishability 
has been given by Ghirardi et al.;\cite{gnrw} 
our preferred interpretation, and a treatment of indistinguishability that
does not jeopardize locality, can be found in reference \cite{ak}. 
Whether these models (or later theoretical developments along similar
lines) are relevant to physics will ultimately be decided by
experiment. 
Theoretical and experimental programs investigating
the physics that might underlie state vector reduction seem to us very strongly
motivated; one purpose of the present article is to argue against
the view --- more often stated than explained --- 
that the problems these programs seek to investigate have already
been solved.  

\section{Acknowledgements}

It is a pleasure to thank S. de Alwis, J. Anandan, W. Boucher, S. Carlip, 
P. Goddard, J. Halliwell, R. Penrose, R. Wald, F. Wilczek 
for interesting discussions and explanations of their views on various
many-worlds interpretations.
Thanks also to a referee for constructive criticisms, and for
permission to quote from these. 
The hospitality of the Institute for Advanced Study is gratefully 
acknowledged. This work was supported by DOE grant DE-AC02-76ER02220.

\end{document}